\documentclass[5p,article]{elsarticle}

\usepackage{lineno,hyperref}
\modulolinenumbers[5]
\usepackage{graphicx} 
\usepackage{natbib} 
\usepackage{amsmath} 
\usepackage{amssymb}
\usepackage{tabularx}
\usepackage{url}
\usepackage{doi}
\usepackage{listings}
\usepackage{mathtools}
\usepackage{multirow}
\usepackage{algorithm}
\usepackage[noend]{algpseudocode}

\usepackage{etoolbox}
\journal{Astronomy \& Computing}

\begin{document}

\begin{frontmatter}

\title{A Processing Pipeline for High Volume Pulsar Data Streams}

\author[mymainaddress]{R.~J. Lyon\corref{mycorrespondingauthor}}
\cortext[mycorrespondingauthor]{Corresponding author}
\ead{robert.lyon@manchester.ac.uk}

\author[mymainaddress]{B.~W. Stappers}
\author[mymainaddress]{L.~Levin}
\author[mymainaddress]{M.~B. Mickaliger}
\author[mymainaddress]{A.~Scaife}

\address[mymainaddress]{School of Physics and Astronomy, University of Manchester, Manchester, M13 9PL, UK, Tel.: +44 (0) 161 275 4202}

\begin{abstract}
Pulsar data analysis pipelines have historically been comprised of bespoke software systems, supporting the off-line analysis of data. However modern data acquisition systems are making off-line analyses impractical. They often output multiple simultaneous high volume data streams, significantly increasing data capture rates. This leads to the accumulation of large data volumes, which are prohibitively expensive to retain. To maintain processing capabilities when off-line analysis becomes infeasible due to cost, requires a shift to on-line data processing. This paper makes four contributions facilitating this shift with respect to the search for radio pulsars: i) it characterises for the modern era, the key components of a pulsar search science (not signal processing) pipeline, ii) it examines the feasibility of implementing on-line pulsar search via existing tools, iii) problems preventing an easy transition to on-line search are identified and explained, and finally iv) it provides the design for a new prototype pipeline capable of overcoming such problems. Realised using Commercial off-the-shelf (COTS) software components, the deployable system is open source, simple, scalable, and cheap to produce. It has the potential to achieve pulsar search design requirements for the Square Kilometre Array (SKA), illustrated via testing under simulated SKA loads.
\end{abstract}

\begin{keyword}
pulsars: general \sep methods: data analysis \sep methods: statistical \sep techniques: miscellaneous
\end{keyword}

\end{frontmatter}


\section{Introduction}
\label{sec:intro}
State-of-the-art pulsar search pipelines are comprised of two principal components. The first is a signal processing system. It converts the voltages induced in the receiving element of a radio telescope into digital signals, and identifies those `significant' detections rising above the noise background. The signal processing system corrects for phenomena such as signal dispersion, excises radio frequency interference (RFI), and optimises search parameters yielding higher signal-to-noise ratio (S/N) detections. The signal processor ultimately produces some number of pulsar `candidates' for analysis. These are time and frequency averaged data products describing each detection. For a complete description of the search process refer to \cite{Lorimer:2005:vm}.\newline

The second search component is a filtering system. It identifies those candidates most likely arising from legitimate astrophysical phenomena, background noise, or terrestrial RFI. In principal this system allows signals of legitimate scientific interest to be isolated and set aside. For clarity we refer to the first component as the signal processor (SP), and the second as the data processor (DP). This paper focuses on the development of a new data processor. In particular we contribute the first realisation of a Commercial off-the-shelf (COTS) based DP for pulsar data designed to operate within an incremental data streams (i.e. `tuple-at-a-time' or `click-stream'). This represents an advancement over the current state-of-the-art, and a departure from traditional off-line batch methods.\newline

Our prototype pipeline is comprised of many software components, some of which are described and experimentally evaluated elsewhere. This applies to the machine learning (ML) techniques deployed in this work in particular \citep{Lyon:2014:jk,Lyon:2016:bs}. Where such work has already been evaluated against the state-of-the-art, we do not re-evaluate here (e.g. see section 6.4 for further justification).\newline

Whilst our work arises from the domain of radio astronomy, our processing scenario is closely related to real-time classification/filtering more generally. Our pulsar candidate data processor is analogous to, for example, the filtering systems employed by the Atlas experiment to process particle detections \citep{Nakahama:2015:yn}, the tools used to identify fraudulent transactions in financial data \citep{West:2016:jw}, and the software used to detect security intrusions in computing systems \citep{HungJen:2013:hj}. This work therefore has wide applicability.
\subsection{Related Work: Pulsar Data Processors}
Modern DPs are comprised of custom software tools, written by research groups in the radio astronomy community \citep[e.g.][]{vanstraten:2011:dspsr,sigproc,PulsarHunter,presto}. These evolve over time, accommodating new algorithmic advances as appropriate. It is not uncommon for such advances to result in the discovery of important, previously unknown phenomena. Where existing tools are perceived to fall short, new ones emerge \citep[e.g.][]{Smith:2016:km,Bassa:2017:cb,Zackay:2017:bz}. Emerging tools are often tailored to achieve specific science goals (e.g. improve sensitivity to longer period pulsars), and are often written with specific observing set-ups, or instruments in mind \citep[issues reviewed and described in][]{LyonPhD:1,Lyon:2016:bs}. There are multiple DP systems in use at any one time, spanning global pulsar search efforts.\newline

Almost all existing DP back-ends execute tasks sequentially, on batches of observational data \citep{LyonPhD:1,Lyon:2016:bs}. The batches are usually processed off-line \citep{Sclocco:2015:be}, either upon completion of a survey, or at the end of an observational session \citep[e.g.][]{Allen:2013:ks,Stovall:2013:ks,Coenen:2014:tv,Bhattacharyya:2016:sc}. A small number of real-time pipelines have appeared. These are designed to detect transient events \cite{McLaughlin:2009:mm,Thompson:2011:kw}, fast radio bursts (FRBs) \cite{Lorimer:2007:mm,Keane:2016:hg,Law:2014:gc,Chennamangalam:2015:ab,Petroff:2015:mb,Karastergiou:2015:ac}, pulsars \citep{Naidu:2015:cb}, or a variety of phenomena \citep{Barr:2014:Superb,Cranmer:2017:bf,Chennamangalam:2017:ab}. The adoption of real-time pipelines has occurred due to changing science requirements (desire/necessity for rapid follow-up), and in response to data storage pressures. These pressures are characterised by increasing data capture rates, which in turn yield increasing volumes of data \citep{Lyon:2013:jk,Lyon:2014:jk,Lyon:2016:bs,Sclocco:2015:be}. Data pressures make it increasingly difficult to process data off-line, primarily as it becomes too costly to buy the media needed to store the data \citep{Sclocco:2015:be,Lyon:2016:bs}. Such challenges are now impacting the search for pulsars. Without the capacity to continually store new data, search methods must be redesigned to maintain their effectiveness as the transition to real-time processing occurs \citep[][]{Barsdell:2012:br,De:2016:gupta,Nieuwpoort:2016:rfi,Cranmer:2017:bf}.\newline
\subsection{Our Contributions}
This paper makes four contributions in this area. Sections 2 and 3 introduce the core research problem, and describe the standard components of an off-line search pipeline (first contribution). After considering the feasibility of transitioning to on-line pulsar search (second contribution), and setting out the associated problems (third contribution) Sections 4-5 consider software tools useful for building a new on-line pipeline. Section 6 goes further, and presents a prototype pipeline design (final contribution). Sections 7-9 evaluate the prototype, and present results describing its performance. Finally Section 10 concludes the paper, and reviews the work.
\begin{figure*}[htp]
	\centering
		\includegraphics[scale=0.5]{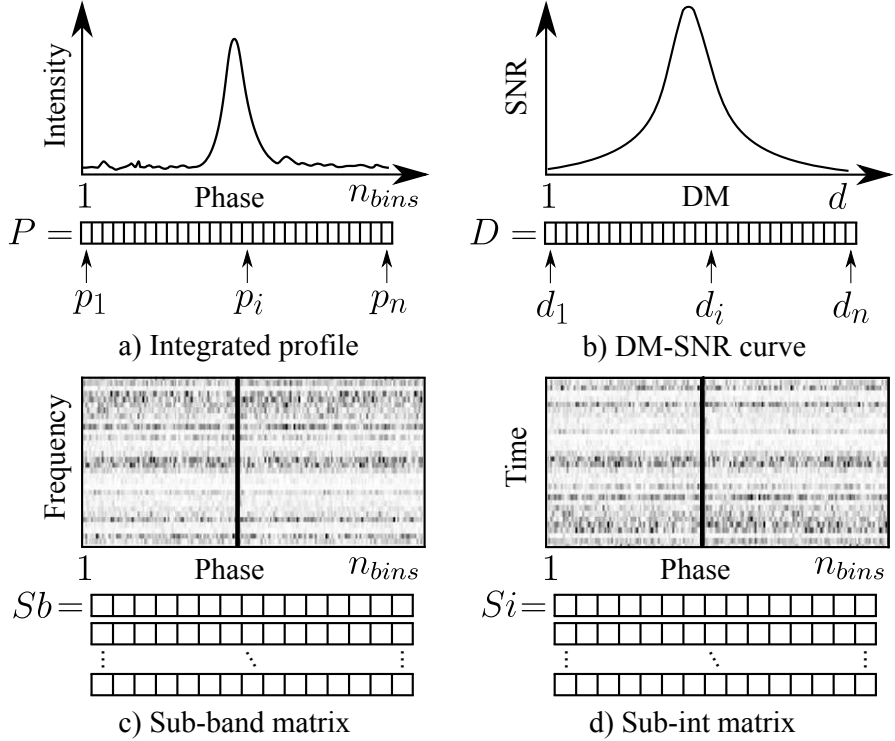}
		\caption[]{Diagram showing the components of the standard pulsar candidate. Plot a) shows the integrated pulse profile, plotted from the vector $P$. Plot b) shows the DM-SNR curve obtained from the vector $D$. Plot c) shows the sub-band matrix, describing the persistence of the signal in frequency. The sub-integration matrix in d) is similar, expect it describes the persistence of the signal in time.}
	\label{Fig:candChars}
\end{figure*}
\section{Off-line Problem Definition}
The goal for a DP is to reduce a set of candidate detections $C$, to a subset of promising candidates $C^\prime$. These are the detections expected to possess the most scientific utility. The input set contains $N$ elements, and $N$ varies according to the exact processing configuration used\footnote{For one observation $N \geq 1000$, for a survey $N > 10^{6}$ is normal.}. The DP should return a set $C^\prime$, such that $|C^\prime| \leq |C|$, though we desire $|C^{\prime}| \ll |C|$.\newline

Each element in $C$ is describable as a candidate tuple. A tuple is a list of $m$ elements. An individual tuple is defined as $c_{i} = \lbrace c_{i}^{1}, \ldots , c_{i}^{m}\rbrace$. Here each element is uniquely identifiable in $C$ via the index $i$. For all $c_{i} \in C$, it holds that $|c_{i}| > 0$. For simplicity all $c_{i}^{j} \in \mathbb{R}$, and there is no implicit ordering in $C$. In practice the numerical components of a tuple represent a detection, or its derived characteristics. The standard components usually stored within $c_{i}$, are shown in Figure \ref{Fig:candChars}. The integrated pulse profile shown in a), is an array of continuous variables describing a longitude-resolved version of the signal averaged in time and frequency. The DM curve shows the relationship between candidate S/N and the dispersion measure (DM) \cite{Lorimer:2005:vm}. Persistence of the detection throughout the time and frequency domains is shown in c) and d).\newline

Persistence in frequency is represented by a two-dimensional matrix c), showing pulse profiles integrated in time for a set of averaged frequency channels (i.e. not full frequency resolution). Persistence through time is represented by a two-dimensional matrix d), showing the pulse profile integrated across similarly averaged frequency channels as a function of time. Other characteristics are also recorded. These can include the S/N, the DM, pulse period, pulse width, beam number, and acceleration; though other metrics are used \citep{LyonPhD:1}. The DP uses these characteristics to make accurate filtering decisions, producing as pure an output set as possible.\newline

We note that $C$ is not a multi-set, i.e., it does not contain exact duplicates. There are however non-exact duplicates present. These include pulsars detected by different telescope beams, or at slightly different characteristic values. Only the `strongest' detection need be retained, with sub-optimal detections discarded to minimise $|C^{\prime}|$. The strongest detection usually has the highest S/N. When $C$ is available off-line, the strongest detection is found via exhaustive comparisons in the worst case.
\subsection{Existing Data Processors}
Existing pipelines execute filtering tasks sequentially upon $C$. The data is processed either after the completion of a pulsar survey, or at the end of an individual observational session if a faster pace of discovery is desired \citep[see for example,][]{Barr:2013:dj,Ng:2012:cn,Coenen:2014:tv,Bhattacharyya:2016:sc}. At present it is feasible to permanently store all candidates in $C$. This allows detections to be processed more than once. It is common for new pulsars to be found in $C$, even after it has been searched multiple times \citep[e.g.][]{Keith:2009:jo,Mickaliger:2012:dl}. This happens when improved search algorithms/parameters are applied, revealing previously hidden detections.\newline

There are some components common to all pulsar DP pipelines. These include `sifting', known source matching, feature extraction, candidate classification and candidate selection. Together these successively filter candidates producing a purer set for analysis. The key components are described in the sections that follow, so their function and computational requirements are understood.
\subsubsection{Sifting}
Sifting is an off-line matching problem. The goal is to to accurately identify duplicate detections in a candidate data set (i.e. find harmonically related duplicates), via comparing candidate pairs. For each arbitrary pair $c_{i}$ and $c_{k}$, where $i\neq k$, a na\"{i}ve sift will exhaustively compare all possible pairs using a similarity measure $s$. The measure is normally associated with a decision threshold $t$, applied over one or more variables in a candidate tuple. If $s$ is above some threshold, the pairing is considered a match. Otherwise the pair is considered disjoint.\newline

The accuracy of the similarity measure can be quantified, when the ground truth matching is known. In such cases the performance of $s$ can be measured using a simple metric $p$, such as accuracy of the output matching,
\begin{eqnarray}
\textrm{Matching Accuracy} = \frac{\textrm{Total matched correctly}}{|C^{\prime} |}\textrm{.}
\end{eqnarray}

The current `best' sift implementation \citep[e.g. in the Seek tool,][]{sigproc} performs an optimised comparison of each $c_{i}$, to every $c_{k}$ in $C$. This has a memory complexity of $\mathcal{O}(n)$ as all $n$ candidates must be stored in memory. The corresponding run time is approximately $\mathcal{O}(n^{2})$. We note that minor adjustments to the approach can yield better runtime performance. Using combinatorics we find that a decreasing number of comparisons only need be done for each $c_{i}$. The total number of permutations for a set of length $n$, where $k$ items are compared at a time, is given by the binomial coefficient,
\begin{eqnarray}
\frac{n!}{(n-k)!\cdot k!}\textrm{.}
\end{eqnarray}
This approach is dominated by $\mathcal{O}(n!)$ for all $k$. In practice for the required $k=2$ the runtime is dominated by $\mathcal{O}(\frac{1}{2}(n-1)n)$, an improvement over the worst case.
\subsubsection{Feature Extraction}\label{subsubsec:features}
Numerical variables known as `features' are usually extracted from candidates post-sifting. These are useful for deriving accurate filtering decisions, and are stored within each candidate tuple. Historically, features were comprised of standard signal characteristics (DM, pulse width, etc) used by human experts to filter candidates manually. Today features are utilised principally by ML algorithms \citep[see][]{Duda:2000:hp,Bishop:2006:pr,Russell:2009:pn}. These build mathematical models able to filter and separate candidate data automatically, based on their feature values. For pipelines employing ML tools, features are generally more complicated than simple signal characteristics (see Section 2.1.3 for more details). The computational cost of extracting features varies. The costs are rarely reported, though can be very high \citep{LyonPhD:1}.
\subsubsection{Candidate Selection}
Candidate selection applies filtering decisions using the features that comprise a candidate tuple. Candidates passing through this stage are `selected' for further processing or study. Candidate selection varies in complexity. From a single threshold applied over S/Ns, to far more complex automated machine learning-based approaches \citep[e.g][]{Eatough:2010:uz,Bates:2012:mb,Zhu:2014:ab,Morello:2014:eb,Lyon:2016:bs}. Where only thresholds are applied, runtime and memory complexity is constant per candidate. For machine learning and other more sophisticated methods, complexities are hard to determine as they are input data dependent. Candidate selection is notoriously difficult. In recent years a rise in candidate volumes has spawned what has become known as the `candidate selection problem' \citep{Eatough:2010:uz}.
\subsubsection{Known Source Matching}
Known source matching involves determining which detections correspond to known pulsar sources. The procedure is often carried out during sifting, though can be done independently after candidate selection. It requires a set $K$ of known sources, usually obtained from a known pulsar catalogue \citep[such as,][]{atnf}. For simplicity\footnote{Known sources are well studied, thus more information is available describing them, than for candidates.} each source in $K$ is defined as $k_{i} = \lbrace k_{i}^{1}, \ldots , k_{i}^{m}\rbrace$, with each tuple uniquely identifiable in $K$ via the index $i$. As before for candidates, all $k_{i}^{j} \in \mathbb{R}$, with the meaning of each $k_{i}^{j}$ the same as for candidates (or at least mappable). A brute force matching approach compares each candidate $c_{i}$ to every $k_{i} \in K$. This corresponds to a runtime complexity of $\mathcal{O}(n \cdot |K|)$. As new pulsars (and other possible radio sources) continue to be found over time, $|K|$ is gradually increasing. The brute force method is thus computationally expensive\footnote{It does not reach quadratic complexity, as $|K| \ll n$.} for an increasing $|K|$ and an unbounded $n$. The memory complexity is $\mathcal{O}(|K|)$, as each known source is typically stored in memory to facilitate fast matching.

\subsubsection{Manual Analysis}
Following the previous steps, a set of candidates $C^{\prime}$ will be stored for manual analysis. Here experts manually examine the elements of $C^{\prime}$, and judge their discovery potential. Given the large number of candidates usually in $C^{\prime}$, this process is time consuming. This in turn introduces the possibility for human error \citep{Lyon:2016:bs}. Manual processing steps may have to be re-run when optimised search parameters/improved search methods are applied to $C^{\prime}$.
\subsection{Feasibility of Real-time Search}
Empirical data describe a trend for increasing survey data capture rates over time \citep{Lyon:2016:bs}. This is expected to continue in to the Square Kilometre Array (SKA) era \citep[][]{Smits:2009:dc,Dimoudi:2015:wa,Broekema:2015:pc,LyonPhD:1}. The projected data rates are large enough to make the storage of all raw observational data impossible (e.g. filterbank/voltage data), and the storage of all reduced data products (e.g. time vs. frequency vs. phase data cubes) impractical \citep{Lyon:2016:bs}. Without the capacity to store data for off-line analysis, SKA pulsar searches will have to be done in real-time at SKA-scales \citep{Dimoudi:2015:wa,Sclocco:2015:be,LyonPhD:1}.\newline

To overcome this problem, selection methods must be adapted to the real-time paradigm. A real-time DP must operate within the time and resource constraints imposed by instruments such as the SKA \cite{Dewdney:2013:pe,Dewdney:2015:pe,Nijboer:2015:rb,Tan:2015:ct}. DP operations must also execute functionally within those constraints. This is difficult to achieve. Not all DP operations, as currently deployed, can run within a real-time environment. The following problems reduce the feasibility of transitioning off-line pipelines to the on-line paradigm:
\begin{itemize}
\item Sifting is an inherently off-line operation that requires data to be aggregated. This is problematic in a real-time environment. The time taken to aggregate data can violate latency requirements. Whilst the memory required to retain all $N$ candidates may not be available. For observations where rapid follow-up is crucial (fast transients), the delay induced via aggregation delays a rapid response.
\item Pulsar feature extraction has never been undertaken on-line, and is currently an off-line batch process.
\item Known source matching, similar to sifting, is currently executed over all candidates off-line. It therefore suffers from similar issues.
\item The most accurate selection systems in use today, are built using off-line machine learning algorithms. Such off-line systems are inherently insensitive to distributional changes in the input data over time. Yet sensitivity to change is an important characteristic for an on-line system to possess, as it enables automated learning and operation. 
\end{itemize}
 Given such challenges, existing methods must be redesigned, or new techniques developed, to maintain the effectiveness of DPs for future pulsar searches. In some cases converting pipeline tasks to run on-line is deceptively trivial. For instance sifting could be done on-line, via only comparing each $c_{i}$ to $c_{i-1}$ and $c_{i+1}$. Whilst functionally possible, such an approach would significantly reduce sifting accuracy. This would in turn produce many more false positive detections, increase the lead-time to discovery, and possibly preclude some discoveries being made. 
 
\section{On-line Problem Definition}
\begin{figure}
	\centering
		\includegraphics[scale=1.0]{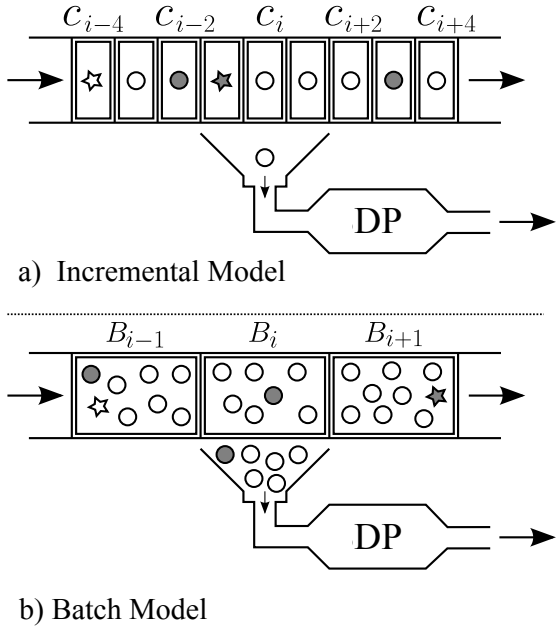}
		\caption[Incremental and batch models of processing.]{Incremental and batch data processing models. The shapes moving through the models represent different item categories (e.g. pulsar vs. non-pulsar). The true class labels are known a priori for the shaded items only.}
	\label{Fig:batchVSinc}
\end{figure}

In a data streaming environment, only a portion of $C$ is available for processing at any given time. Either due to real-time constraints forcing batches of data to be processed separately, or $C$ being too large to process in a single pass. In either case there are two processing models that can be employed \cite[discussed elsewhere e.g.,][]{Zaharia:2013:z,Carbone:2015:ab}. The incremental `tuple-at-a-time' model applies when individual data items arrive at discrete time steps. An item $c_{i}$ arriving at time $i$, is always processed after an item arriving at time $i-1$, and before $i+1$. The batch (or micro-batch/discretized streams) model applies when groups of items arrive together at discrete time steps. Here batch $B_{i}$ arrives after $B_{i-1}$ and before $B_{i+1}$. This is summarised in Figure \ref{Fig:batchVSinc} for clarity. Both models temporally order data, which has implications for how the data can be processed.\newline

The batch model is advantageous when groups of items exhibit distributional similarity. For instance when processing data from seismic sensors, it makes sense to work with batches, as patterns of seismic activity will likely be close temporally. Whilst if looking for fraudulent activity in a stream of random financial transactions, each transaction should be processed in isolation. Otherwise an innocent transaction may be incorrectly linked to a fraudulent one. There is a trade-off between computational efficiency and practical utility \cite[][]{Chintapalli:2016:sc,Lopez:2016:ll}, that must be struck when choosing a processing model.\newline

For the processing of single pulse events, it does not make sense to batch data. Waiting for a batch incurs a time penalty which impedes our ability to initiate rapid follow-up. For pulsar search batch processing is possible, however batch sizes are potentially very large. In both cases it would appear that an incremental model is the simplest to adopt.
\subsection{An Incremental Model for Data Processing}
We extend the notation used previously. The candidate set $C$ is now a candidate stream $C=\lbrace c_{1}, c_{2}, \ldots, c_{n}, \dots \rbrace$. Here each $c_{i}$ is delivered at time step $i$, and $c_{i}$ always arrives before $c_{i+1}$ (discrete time model with temporal ordering). The size of the set $C$ is unbounded in this scenario (unknown $N$). This assumes a limitless supply of individual candidates, redefining the SP as a data stream producer, and the DP as a data stream consumer.\newline

\subsection{Practical Assumptions}
We assume a worst case scenario, where each data item is processed just once, as per the incremental model. Filtering decisions must be made on an individual candidate basis, with limited knowledge. For example, suppose $c_{2}$ arrives for processing where only $c_{1}$ has been observed. Here a filtering decision can only be made using knowledge of the candidate/s already seen and $c_{2}$. If $c_{2}$ is most likely noise or interference, then the decision is conceptually simple. Yet the decision making process becomes complicated in scenarios where multiple detections of the same source are likely to be made during an observation. If $c_{2}$ is indeed a pulsar detection, do we retain it? Is it the strongest detection we are likely to see? If it is weak, should we throw it away, assuming a stronger detection will come along? Do we keep every weak detection, and risk increasing the size of $C^{\prime}$? There are no simple answers to such questions.\newline

To proceed in our domain, we assume candidates will be ordered according to $c_{i}^{0}$ (an ordering variable). This ordering must be strict, so that $\forall c_{i} \in C$, $c_{i}^{0} \leq c_{i+1}^{0}$. This definition specifies an ascending order, though it would make no difference if it were descending. By using an ordering we can have confidence that similar items will be close together in the stream (close temporally). This simplifies our data processing, and will be used to develop an on-line sifting algorithm in Section 6.2. Note that we are implicitly assuming there exists an upstream processing component, capable of ordering data items correctly. We accept that it may not always be possible to assign an ordering conducive to improved data processing. However, where possible we should attempt to order our data in a meaningful way, as it greatly assists with the downstream processing. We now look for frameworks that can accommodate this processing model.
\section{Candidate Frameworks for Prototyping}

The design of a DP is driven by multiple considerations. We have focused upon utilising COTS software components to reduce cost. This is our primary design driver. However our choice of software components was also driven by the need to reduce computational overheads, whilst maximising scalability and design modularity.
\subsection{Review \& Chosen Framework}
Signal processing systems often utilise accelerator hardware to enable real-time operation. This includes Graphics Processing Units (GPUs) \citep[e.g.][]{Magro:2011:ak,Couturier:2013:rc,Dimoudi:2015:wa} and Field Programmable Gate Arrays (FPGAs) \citep[e.g.][]{Kilts:2007:ks,Dubey:2008:dr,Woods:2008:rm,Vanderbauwhede:2013:wb,Haomiao:2015:wo}. In contrast most DP software is executed on general purpose computing resources, with some exceptions\footnote{Machine learning based filters \citep[e.g.][]{Zhu:2014:ab}.}. We therefore only consider DP execution upon standard Central Processing Units (CPUs). There are many frameworks/approaches that meet our criteria \citep[for a review see e.g.,][]{Gorawski:2014:mg}. Yet we chose Apache Storm \citep{Jain:2014:an,storm} to support our new on-line prototype. It was chosen due to its application in the real-world to similar large-scale incremental processing problems \citep[see][]{Jones:2013:twitter,Toshniwal:2014:ta,Weiler:2015:am} and strong user support base. Storm is similar to frameworks such as Apache Hadoop \citep{White:2012:wt,Jankowski:2015:mp,hadoop}, Samza \citep{Kleppmann:2015:mk,samza}, Spark \citep{Zaharia:2014:sp,spark} and S4 \citep{Neumeyer:2010:ln,s4}, and is fairly representative of similar modern COTS tools. Since completing this work, we discovered other frameworks in the software ecosystem similar to Storm. These include Apache Samoa \cite{Bifet:2015:as,samoa}, Apache Flink \cite{Carbone:2015:ab}, and Kafka Streams \cite{Shree:2017ks}. The use of Apache Storm in lieu of these systems, does not undermine the novelty or usefulness of the contributions of this work. Principally as the processing pipeline proposed in Section 6., is framework independent and compatible with any tool capable of supporting the `tuple-at-a-time' processing model. 
\section{Apache Storm-based Processing Framework}
Storm is a Java-based framework. It operates under an incremental data stream model. It is underpinned by the notion of a directed-acyclic graph \citep[DAG, see][]{Thulasiraman:1992:dg}. The graph models both the processing steps to be completed, and the data flow. The graph is known within Storm as a \textit{topology}. Storm topologies only allow data to flow in one direction. This makes recursive/reciprocal processing steps impractical to implement (though newer frameworks discussed in Section 4 can overcome such limitations).
\subsection{Topologies}
A topology is comprised of nodes. Nodes represent either data output sources known as \textit{spouts}, or processing tasks applied to the data called \textit{bolts}. Data flows from spouts to bolts via edges in the topological graph. Individual data items are transmitted in the form of $n$-tuples. These are finite ordered lists much like the candidate tuples defined in Section 2.
\subsection{Edges}
Edges represent generic connections between processing units, made via a local area network (LAN), or a wide area network (WAN). Data flows via the edges as tuples. Upon arriving at a bolt a tuple is usually modified, and the updated tuple emitted for a downstream bolt to process. The practical data rates between edges are non-uniform, and often subject to change. Fluctuating data rates can lead to resource contention, if too much data is funnelled through too few processing bolts.
\subsection{Bolts}
Each bolt encapsulates a modular computing task, completable without interaction with other bolts. Such modularity allows data to be processed in parallel, across heterogeneous computing resources without a loss in functionality. The duplication of bolts allows the topology to be scaled to increasing amounts of data, without additional development effort. It also means that when a bolt fails, it can quickly be replaced with a new instance. The modular design is not ideal for all scenarios. Particularly those where the persistence of state is required. Generally state persists within a spout/bolt only whilst it is executing. If it fails or is stopped, that state is lost. Storm more generally is unsuitable for processing components that are i) not lightweight, ii) require a persisted state to function correctly, or iii) require data to be aggregated/batched.
\subsection{Tuple Flow}
It is possible to control the flow of tuples through a topology. This is achieved via a cardinality relation. This is defined between spouts and bolts via edges. The relation can be one-to-one, one-to-many, or one-to-all. If using a one-to-one relation, tuples either move forward to a single random bolt, or a value-specific bolt based on a tuple attribute test\footnote{For example if $c_{i}^{1} \leq 10$, send to bolt $b_{1}^{1}$, else to bolt $b_{1}^{2}$.}. Upon transmission of a tuple, a `tuple received' acknowledgement can be requested by the sender (originating spout or bolt). If the sender receives no acknowledgement, it will resend the tuple. This enables Storm to maintain the property that all data is processed \textit{at least} once, helping ensure fault tolerance.
\subsection{Deployment}
Topologies execute upon clusters comprised of two distinct nodes. There are i) master nodes that run a daemon called `Nimbus', and ii) one or more worker nodes running a daemon called `Supervisor'. A master node is responsible for executing tasks on the worker nodes, and restarting spouts/bolts after failures. The worker nodes execute spouts and bolts separately, within threads spawned by a Java Virtual Machine\footnote{One thread per worker node's physical CPU core.} (JVM).\newline

 Communication between the master and worker nodes is coordinated by Apache Zookeeper \citep{Jankowski:2015:mp}. Zookeeper maintains any state required by the master and the supervisors, and restores that state in the event of failure. A Storm cluster therefore consists of three components: Nimbus, one or more Supervisors, and a Zookeeper.
\begin{figure*}[htp]
	\centering
		\includegraphics[scale=0.73]{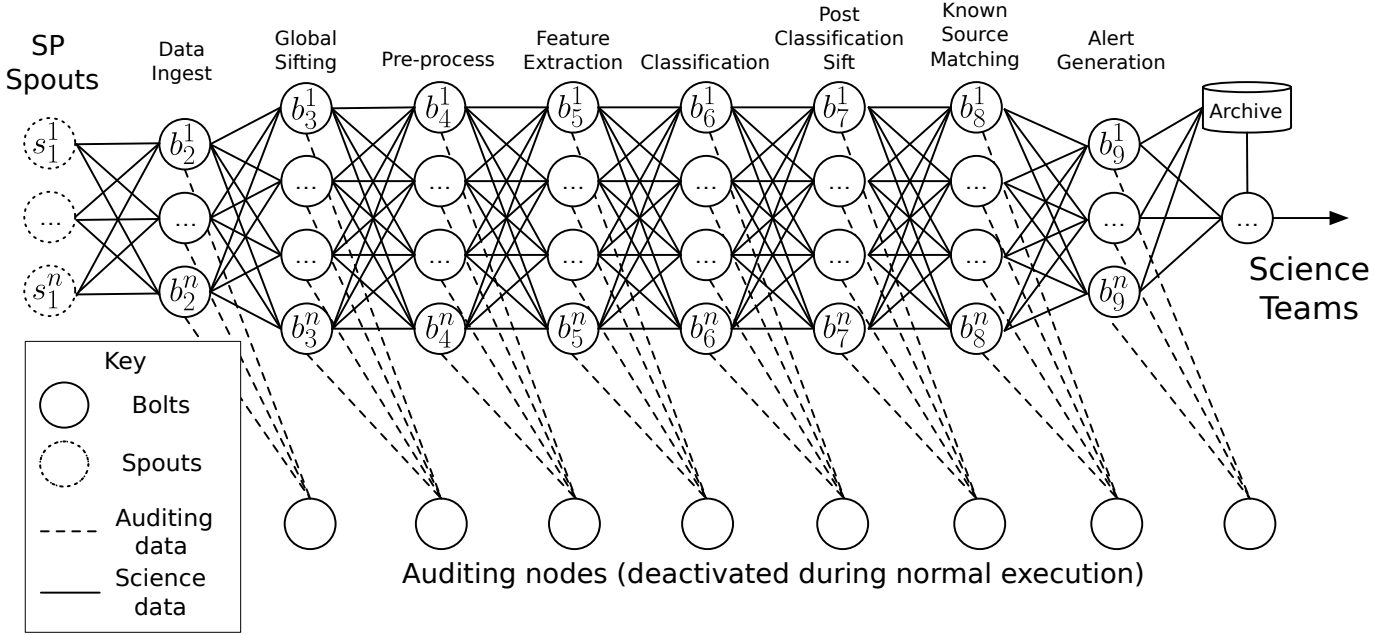}
		\caption[]{The prototype DP topology. Data is supplied to the topology via SP spouts, which generate candidate tuples. The tuples are propagated forwards through the topology. No tuple is duplicated, and each reaches only one bolt in the topology at any given time. There are nine distinct types of bolt in the topology, described in the sections that follow. }
	\label{Fig:prototpyeTopology}
\end{figure*}
\section{DP Prototype Design}
The prototype design is shown in Figure \ref{Fig:prototpyeTopology}. It has a single `layer' of input spouts, which emit candidate tuples at a controllable rate. Candidates emitted from the spouts are propagated to a single random bolt, at each subsequent layer in the topology. The first layer contains data ingest bolts. These modify the data so that it is transmitted in a format amenable to further processing.\newline

The second layer attempts to filter out duplicate candidates via sifting. To achieve this, the sifting bolts employ a new distributed global sift algorithm. This is described in more detail in Section \ref{subsec:online_sifting}. After sifting, tuples are sent to bolts that perform data pre-processing (i.e. normalisation). This is required prior to ML feature extraction. Once complete, the next layer of bolts extracts the features. The features are appended to the tuple, and sent onwards for further processing.\newline

Following feature extraction, tuples are passed to ML classification bolts. These execute an on-line ML algorithm developed by \cite{Lyon:2016:bs}. This predicts the true class origin of each candidate. The predictions are appended to each tuple, and passed on to secondary sifting bolts. These remove duplicates in lieu of the additional information obtained during ML classification. Tuples making it through the second sift, are passed to source matching bolts. These attempt to match promising candidates to known pulsar sources in a pulsar catalogue. Candidates not matched to known sources (likely new pulsars) generate alerts for follow-up action.\newline

The topology has been designed to provide auditing capabilities. Auditing nodes are connected to the processing bolts, via the dashed lines shown in Figure \ref{Fig:prototpyeTopology}. They are used to audit the performance of individual bolts, or the entire topology during testing. The auditing nodes are not activated during normal execution.\newline

The individual spouts/bolts are now described in more detail. Source code for the procedures described in the following sections can be found on-line \citep{Lyon:2017:psm}. The code is provided in the form of an interactive iPython notebook. It contains additional details that support the information presented in this paper.
\subsection{SP Spouts}
At present no SP system can generate data fast enough to stress our prototype. We therefore created SP spouts which emit real candidate tuples driving the data processing. These play the role of the SP, delivering data at a controllable rate. The spouts are highly customizable. They can generate a fixed number of candidates, or as many as possible within a fixed period of time. Importantly the ground truth label for each and every generated tuple is retained. This allows candidate filtering accuracy to be evaluated anywhere within the topology.\newline

To be realistic, the spouts must output real-world candidate class distributions. For the prototype we define a ratio used to achieve this. Suppose each candidate $c_{i}$ is associated with a label defining its true origin. This can be modelled numerically via a binary variable $y$, where $y_{i}\in Y=\lbrace -1,1 \rbrace$. Here $y_{i}=-1$ equates to non-pulsar and $y_{i}=1$ to pulsar. The set of candidates which describe real pulsar detections $P \subset C$ contains only those candidates for which $y_{i}=1$. Similarly $\neg P \subset C$ contains only those candidates for which $y_{i}=-1$. Thus the ratio,
\begin{eqnarray}\label{eq:imbalance_Ratio}
c_{\rm ratio} & = & \frac{| P |}{| \neg P |} \textrm{,}
\end{eqnarray}
describes the imbalance between pulsar and non-pulsar candidates in generated data. The ratio observed during real pulsar searches varies from 1:7,500 \citep{Keith:2010:bl,ThorntonPhD:1} to 1:33,000 \citep{Cordes:2006:kh,Lazarus:2012:palfa,PAlfa:2015:pf}. For the prototype, we maintain ratios of up to ~1:10,000 (0.0001). Evidence suggests this figure to be representative \citep{Lyon:2016:bs} and challenging to deal with.\newline

To maintain the ratio, spouts have been designed to emit two types of candidate. Type 1 candidates are contrived randomly generated non-pulsar tuples. Whilst valid, they simply resemble white noise. Type 2 candidates are representations of real candidates. These were sampled from data obtained during the High Time Resolution Universe Survey South \citep[HTRU, ][]{Keith:2010:bl,Lyon:2015:htru2}. Type 2 candidates provide a genuine test of our prototype's discriminative capabilities. Type 2 candidates can describe either the pulsar or non-pulsar class. As the sample of type 2 candidates is small, some pass through the pipeline more than once during testing. This does not invalidate our results, as duplicates must be passed through the system to test our sifting approach.
\subsubsection{Spout Data Rate and Volume }
Input data rates are determined by the length of an observation $t_{\rm obs}$ (seconds), the number of telescope beams $n_{\rm beam}$, the number of candidates returned by the SP per beam $c_{\rm beam}$, and the size of each candidate $c_{\rm size}$ (MB). The total data volume per observation is given by, 
\begin{table}[]
\centering
\label{tab:configuration}
\begin{tabular}{|l|l|}
\hline
Variable            & Value \\
\hline \hline
$t_{\rm obs}$       & 600 seconds \\
$n_{\rm beam}$      & 1,500 (SKA-Mid)       \\
$c_{\rm beam}$      & 1,000       \\
$c_{\rm obs}$       & 1,500,000   \\
$c_{\rm size}$      & ~2.2 MB     \\
$c_{\rm ratio}$      & 0.0001      \\
$d_{\rm rate}$      & ~5.5 GB/s    \\
$c_{\rm rate}$      & 2,500 (per second)       \\
$d_{\rm volume}$    & ~3.3 TB     \\
\hline
\end{tabular}
\caption{Summary of assumptions made when designing the prototype. Here $t_{\rm obs}$ is the observation time in seconds, $n_{\rm beam}$ the number of beams used per observation, $c_{\rm beam}$ the total number of candidates anticipated per beam, $c_{\rm obs}$ the total number of candidates anticipated per observation, $c_{\rm size}$ the size of an individual candidate, $c_{\rm ratio}$ the ratio of pulsar to non-pulsar candidates in input data, $d_{\rm rate}$ the anticipated data rate per second, and finally $d_{\rm volume}$ is the expected total data volume per observation. }
\label{tab:config}
\end{table}
\begin{eqnarray}
d_{\rm volume} = n_{\rm beam} \times c_{\rm beam} \times c_{\rm size} \textrm{.}
\end{eqnarray}
Whilst the SP to DP data rate, assuming a steady uniform transmission, is given by, 
\begin{eqnarray}
d_{\rm rate} = \frac{d_{\rm volume}}{t_{\rm obs}}\textrm{.}
\end{eqnarray}

The assumed parameter values are given in Table \ref{tab:config}. These were chosen after studying SKA design documentation \citep[see][]{Dewdney:2013:pe,Dewdney:2015:pe,Broekema:2015:pc}. Here candidate tuples are comprised of,
\begin{itemize}
\item a 262,144 sample data cube (128 bins, 64 channels, 32 sub-ints), where each sample is 8 bytes in size. This describes the detection in time, phase, and frequency. 
\item the variables; right ascension (RA), declination (DEC), S/N, period, DM, beam number. These variables require 56 bytes of storage space.
\end{itemize}
Together these components produce a tuple $\approx 2.2$ MB in size. For 1,500 beams, and 1,000 candidates per beam, this equates to 1.5 million candidates. The corresponding total data volume is $\approx 3.3$ TB per observation. The $t_{\rm obs}=600$ second time constraint, implies a data rate of 5.5 GB/s. The SP spouts will generate data at and above this rate, when simulating pulsar search operations.
\subsection{On-line Global Sifting}\label{subsec:online_sifting}
Standard sift only compares candidates detected within the same telescope beam \citep[e.g.][]{KarakoArgaman:2015:ac}. Our approach works globally across beams, in principle yielding better results. It sifts one candidate at a time, as opposed to candidate batches. It's success is predicated on two assumptions. First, it assumes a stream partitioning that ensures candidates with similar periods from all beams (measured in $\mu$s) arrive at the same bolts. Thus an individual bolt represents a period range, and together they cover the plausible range of pulse periods. Second, it assumes candidates are ordered (see Section 3.2) according to their pulse periods. This facilitates the partitioning. Since similar period candidates always arrive at the same bolts, it becomes possible to check for duplicates via counting observed periods. The logic underpinning this approach is simple. If a period is observed many times, it is a possible duplicate. When possible duplicates are compared via other variables (e.g. DM) for additional rigour, counting can be used to efficiently find duplicates.\newline

The general approach is summarised in Algorithm 1. This describes the code at a single bolt. Note the algorithm requires sufficient bins to count accurately. As pulsar periods are known to be as low as 1.396 milliseconds \citep{fastpulsar}, and given that pulse periods can be similar at the millisecond level; there must be enough bins to count at microsecond resolution. The algorithm uses the array $F$ to maintain the count, by mapping the array indexes $0,\ldots,n$ to specific periods/period ranges. We use a scaling and a rounding operation to achieve this. Together these find the correct bin index to increment for an arbitrary period. The variables $floor$ and $ceiling$ help to accomplish this. The scaling of the data is done on line 8, with the integer rounding operation done on line 9 which obtains the index.\newline

\begin{algorithm}
\small
\caption{Global Sift}
\begin{algorithmic}[1]
\Require An input stream $	C=\lbrace ..., (c_{\rm i}),...\rbrace$, such that each $c_{\rm i}$ is a candidate, and $c_{\rm i}^{\rm j}$ its $j$-th feature. Here $c_{\rm i}^{\rm 0}$ is the pulse period (ordering variable). Requires a similarity function $s$ which can be user defined, the number of period bins $p_{b}$ to use, the smallest period value expected at the bolt $min$, and the largest period value expected at the bolt $max$.
\Procedure {Global Sift}{$C,s,p_{b},min,max$}

\If{ F = null }
\State $F \leftarrow array[p_{b}]$ \Comment Init. counting array
\State $n \leftarrow 0$ \Comment Init. candidate count
\State $floor \leftarrow 0.0$
\State $ceil \leftarrow p_{b}$
\EndIf
\State $n \leftarrow n+1$ \Comment increment observed count
\State $p \leftarrow ((ceil - floor) * (c_{\rm i}^{\rm 0} - min) / (max - min)) + floor$
\State $index \leftarrow (int)p$ \Comment Cast to int value

\If{ $F[index] > 0$ }
\State F[index]++;
\State \Return s(true); \Comment Similarity check, period seen
\Else
\State F[index]++;
\State \Return s(false); \Comment Similarity check, period not seen
\EndIf

\EndProcedure
\end{algorithmic}
\end{algorithm}
 
 The if-else statement on lines 10-15 contains the only logic in the algorithm. It checks if the period counter is greater than zero (indicating it has already been observed). If $F[index] > 0$, then we pass this information to the similarity function for checking. The algorithm can be modified for improved accuracy. For example, if frequency counts are maintained for multiple variables (DM, pulse width, beam etc.), the probability of a match can be estimated across all of them. Note the similarity function $s$ used on lines 12 and 15, can also be made to check matches counted in neighbouring bins. Candidates counted in bins neighbouring $F[index]$ have very similar periods, and thus should be considered possible duplicates. It is trivial to implement, though requires additional memory and a sliding window of examples. The use of a sliding window is discussed briefly below.\newline

The general approach described has weaknesses. The first candidates arriving at a sift bolt will never be flagged as duplicates, due to no prior periods being been observed. Furthermore, the approach will treat duplicates with different S/Ns as equivalent. However we wish to retain only the highest S/N detection, and discard the rest. The model described thus far cannot achieve this. It cannot anticipate if, or when, a higher S/N version of a candidate will enter the topology. Such weaknesses are a result of the trade-off between computational efficiency, and sifting accuracy. This problem can however be overcome, using a sliding `similarity' window over the data \citep[see ][]{Lyon:2017:psm}. 
\subsection{On-line Feature Extraction}\label{subsubsec:onlinefeatures}
Eight features are extracted from each tuple moving through the topology. These are described in Table 4. of \cite{Lyon:2016:bs}. The first four are statistics obtained from the integrated pulse profile (folded profile) shown in plot a) of Figure \ref{Fig:candChars}. The remaining four similarly obtained from the DM-SNR curve shown in plot b) of Figure \ref{Fig:candChars}.\newline

The computational cost of generating the features in Floating-point operations (FLOP) is extremely low. If $n$ represents the number of bins in the integrated profile and DM curve, the cost is as follows:
\begin{itemize}
\item \textbf{Mean cost:} 2(n-1) + 4 FLOP.
\item \textbf{Standard deviation cost:} 3(n-1) + 9 FLOP.
\item \textbf{Skew cost:} 3(n-1) + 7 floating point FLOP.
\item \textbf{Kurtosis cost:} 3(n-1) + 7 floating point FLOP.
\end{itemize}
This assumes the following: addition, subtraction and multiplication all require 1 FLOP, whilst division and square root calculations require 4. The Standard deviation calculation assumes the mean has already been calculated first. Likewise, the skew and kurtosis calculations assume the mean and standard deviations have already been computed, and are simply reused. The features cost 2,848 FLOP per candidate \citep{LyonPhD:1}. The runtime complexity of the feature generation code is $\mathcal{O}(n)$ over the input space, and memory complexity is $\mathcal{O}(n)$. Note that in both cases $n$ is small.  

\subsection{On-line Classification}
This prototype is focused on processing incremental data streams in real-time. At present only one classifier has been developed specifically for incremental pulsar data streams - the GH-VFDT \citep{Lyon:2014:jk,Lyon:2016:bs,LyonPhD:1}. All other pulsar classifiers developed in recent years can be characterised as off-line supervised systems. Thus in terms of classifiers, the GH-VFDT is the logical choice for this work. Despite this we acknowledge that we may have overlooked classifiers capable of achieving better recall on our data. As the goal of this paper is to illustrate the feasibility of processing pulsar data incrementally in real-time, and not to advocate the use of a specific ML algorithm, this is not something we consider to be problematic. Our design does not preclude the use of any other incremental data stream classifier developed now or in the future.\newline

The GH-VFDT classifier was designed for pulsar candidate selection over SKA-scale data streams. It is an on-line algorithm, capable of learning incrementally over time. It is therefore able to adapt to changing data distributions, and incorporate new information as it becomes available. The GH-VFDT is extremely runtime efficient, as it was designed to utilise minimal computational resources.\newline

The algorithm employs tree learning to classify candidates, described via their features \citep[for details of ML classification see,][]{Duda:2000:hp,Bishop:2006:pr,Russell:2009:pn}. Given some ground truth `training data' describing the pulsar and non-pulsar classes, tree learning partitions the data using feature split-point tests \cite[see Figure 8 in][]{Lyon:2016:bs}. Split points are chosen that maximise the separation between the classes. In practice this involves firstly choosing a variable that acts as the best class separator. Once such a separator is found, a numerical threshold `test-point' is found, that yields the greatest class separability. This process repeats recursively, producing a tree-like structure. The branches of the tree form decision paths. The paths can yield highly accurate classification decisions when given high quality training data.\newline

The memory complexity of the algorithm is $\mathcal{O}(lf\cdot 2c)$ (sub-linear in $n$), where $l$ describes the number of nodes in the tree, $f$ the number of candidate features used ($f=8$ for the prototype), and finally $c$ the number of classes (here $c=2$, pulsar \& non-pulsar). The runtime complexity of the algorithm is difficult to quantify, as it is input data dependent. However it is of the order $\mathcal{O}(n)$.
\subsection{On-line Post Classification Sift}
Post-classification sifting aims to remove duplicate detections in lieu of ML predicted class labels. Here two candidates with very similar period and DM values (other variables can be considered), have their predicted class labels compared. If the same, these are considered likely duplicates. In this case, only the strongest detection need be forwarded on. This approach can be improved if a sliding window over tuples is used \citep[see ][]{Lyon:2017:psm}. This allows similar tuples to be compared together in small micro-batches defined by the window. Only after the window moves away from an unmatched tuple does it get forwarded.
\subsection{On-line Known Source Matching}
\begin{figure*}
	\centering
		\includegraphics[scale=0.7]{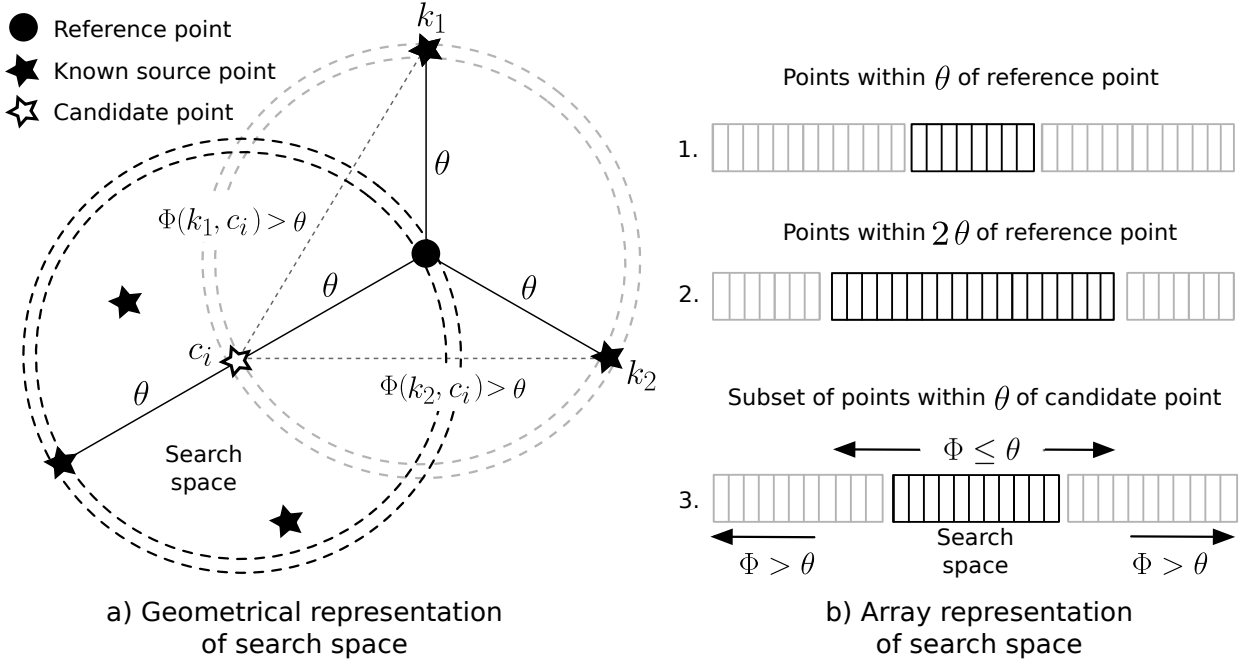}
		\caption[]{A visual representation of how the matching search space is determined. In a) we see candidate source $c_i$, which has an angular separation from the reference point of $\theta$. There are also 2 known sources $k_1$ and $k_2$, with the same separation from the reference point. The angular separation between $c_i$ and other known sources is computed by $\Phi$. If the separation is greater than $\theta$, then $c_i$ is not compared to $k_{i}$. Otherwise a comparison is performed. From this representation it is clear that $c_i$ could be related to known sources up to $2\theta$ from the reference point. However only the region around $c_i$ should be searched. In b) we see the same information, but stored within an array. This is the array used for searching. The sources in the array are sorted according to their angular separation to the reference point. This ordering allows the search region to be found quickly in step 1. This is then refined, producing the narrowed search region in step 3.}
	\label{Fig:Match}
\end{figure*}
We have developed a new, very fast, source matching algorithm. It is built upon a divide and conquer programming approach (built upon tree-search). This involves attempting to recursively divide the matching space, reducing the number of comparisons to be undertaken on each iteration. It relies on an ordering applied over the set of known sources $K$, represented as an array. A total ordering of elements in $K$ is required, according to some variable $k_i^j$. To achieve a total ordering, then for all $k_{i}^{j}$, $k_{i+1}^{j}$, and $k_{m}^{j}$, where $m>i+1$,
\begin{eqnarray}\label{eq:antisym}
\textrm{if } k_{i}^{j} \leq k_{i+1}^{j} \textrm{ and } k_{i+1}^{j} \leq k_{i}^{j} \textrm{ then } k_{i}^{j}=k_{i+1}^{j} \textrm{,}
\end{eqnarray}
\begin{eqnarray}\label{eq:transitivity}
\textrm{if } k_{i}^{j} \leq k_{i+1}^{j} \textrm{ and } k_{i+1}^{j} \leq k_{m}^{j} \textrm{ then } k_{i}^{j}\leq k_{m}^{j} \textrm{,}
\end{eqnarray}
\begin{eqnarray}\label{eq:reflexivity}
k_{i}^{j} = k_{i}^{j} \textrm{,}
\end{eqnarray}
\begin{eqnarray}\label{eq:totality}
k_{i}^{j} \leq k_{i+1}^{j} \textrm{.}
\end{eqnarray}
Here equations 6-9 define the antisymmetry (6), transitivity (7), reflexive (8) and totality properties (9) respectively. To apply the desired ordering, we require a numerical value per source that satisfies these properties. This can be obtained via measuring the angular separation $\theta$, between each known source, and a single reference coordinate ($00^h$ $00^m$ $00^s$ and $00^{\circ}$ $00^\prime$ $00^{\prime\prime}$). This allows sources to be strictly ordered according to their separation from the reference point. This reference value should be computed off-line so that the array $K$ is ordered correctly in advance. Note that here we assume $K$ contains all known sources. This represents a worst case matching scenario. In practice, $K$ will likely be filtered, so that it includes only those known sources in the patch of sky in which the candidates were observed (pre-selected known sources based on pointing location).\newline

 For each candidate source $c_i$ to be matched, the reference separation must also be computed. Intuitively known sources near to $c_i$ would appear to be similarly separated from the reference point. The reality is more nuanced. It is possible for sources to be similarly separated from the reference point, yet be very far apart. This is shown in Figure 7. Here candidate $c_i$ and known source $k_1$, are both separated from the reference point by an angular separation of $\theta$. However these are too distant from one another to be considered a match. This does not affect the matching accuracy of the new approach. The sorting of known sources is only used to reduce the search space. In particular it allows us to find a position from where to start searching in the known source array $K$. We call this position the `search index'. From there, we need only compare $c_i$ to sources in $K$ around the search index, i.e. with separations $\leq 2\theta$ with respect to the reference point. These known sources must be adjacent to the search index given the properties of the ordering.\newline

A divide and conquer algorithm firstly divides up the search space recursively, until a search index is found. This is achieved as follows: if the separation of $c_{i}$ from the reference point, is less than the separation between the reference point and the centre source in $K$, search the first half of $K$. Else, search only the second half of $K$. This process is repeated until the closest single match is found, and a search index returned. Once the closest match is found, a secondary algorithm compares $c_{i}$ to those sources nearest to it. Precisely how many comparisons are done to the left, and right of the search index, is up to the user. For the prototype, comparisons are not undertaken if the distance between $c_{i}$ and $k_{i}$ exceeds $1.5^{\circ}$.\newline

\begin{algorithm}
\small
\caption{Divide \& Conquer Matching}
\begin{algorithmic}[1]
\Require A set of known sources $	K=\lbrace k_{i}, \ldots , k_{m} \rbrace$, the numerical sorting variable $sep$, of the candidate $c_{i}$ to compare. Finally $i_{start}$ is the index to search from in $K$, and $i_{end}$ is the index to search to.
\Procedure {DC}{$i_{start},i_{end}K,sep$}
\If{$i_{end} - i_{start} == 1 $}
\State \Return $i_{start}$  \Comment Only 1 source
\ElsIf{$i_{end} - i_{start} == 2 $}
\State \Return $i_{start}+1$  \Comment 2 sources, arbitrarily pick last
\Else
\State $i_{middle} \leftarrow$ int( ceil( ($i_{end} + i_{start}$) / $2.0$ ) )
\State $k_{i} \leftarrow K[i_{middle}]$ \Comment Get middle source
\If{ $sep< k_{i}.sep$}
\State \Return {DC}{$(i_{start},i_{middle}K,sort)$} \Comment 1st half
\ElsIf{$sep > k_{i}.sep$}
\State \Return {DC}{$(i_{middle},i_{end}K,sep)$} \Comment 2nd half
\Else
\State \Return $i_{middle}$ \Comment Found search location.
\EndIf
\EndIf
\EndProcedure
\end{algorithmic}
\end{algorithm}

The divide and conquer approach is presented in Algorithm 2, and the matching procedure defined in Algorithm 3. The matching procedure compares the period and DM of a promising $c_{i}$, to some potential match $k_{i}$. The known source $k_{i}$ is considered a possible match for $c_{i}$, only if their period and DM values are similar to within a user specified error margin $e_{margin} \in [0,1]$. For example, an $e_{margin}=0.1$ corresponds to a 10\% error margin. When using this margin we consider a known source to be possible match, only if its period and DM are within 10\% of the candidate's ($\pm 5$\%). A 10\% error margin over pulse period is large. A smaller margin should be used in practice, especially for data recorded with high time resolution. However for our test data, pulsar periods differ by up to $\pm 5$\% compared to values recorded in the pulsar catalogue. The prototype does allow for a different margin to be chosen.\newline

The computational complexity of the overall approach is $\mathcal{O}(n \cdot \tau)$, where $\tau$ is a proxy for the number of comparisons made between known sources and candidates based on $\theta$. The modified runtime is practically speaking linear in $n$ (as $\tau$ is usually small). The complete algorithm has been described in an iPython notebook \citep{Lyon:2017:psm}. The approach presented here is inherently simple and fast. It differs from existing cross matching methods used in astronomy, which employ Bayesian methods \citep{Budavari:2008:pc}, or multiple catalogues describing sources at different wavelengths \citep{Fan:2015:df}.
\begin{algorithm}
\small
\caption{Matching Procedure}
\begin{algorithmic}[1]
\Require A known source $k_{i}$, a candidate $c_{i}$, an angular separation used for matching $\theta$, and an accuracy level for period and DM matching $e_{margin} \in [0,1]$.
\Procedure {isMatch}{$k_{i},c_{i},\theta,e_{margin}$}
\State $c_{p} \leftarrow c_{i}$ \Comment Get period from candidate
\State $c_{dm} \leftarrow c_{i}$ \Comment Get DM from candidate
\State $k_{p} \leftarrow c_{i}$ \Comment Get period from known source
\State $k_{dm} \leftarrow c_{i}$ \Comment Get DM from known source
\State $p_{diff} \leftarrow (e_{margin} \times c_{p}) / 2$
\State $dm_{diff} \leftarrow (e_{margin} \times c_{dm}) / 2$
\State $hms \leftarrow  [1, 0.5, \ldots ,0.03125]$ \Comment Harmonics to check
\For{$h \leftarrow 0$, $h{+}{+}$, while $h < |hms|$}
\If{$c_{p} > (k_{p} *hms[h]) - p_{diff}$}
\If{$c_{p} < (k_{p} *hms[h]) + p_{diff}$}
\If{$c_{dm} < k_{dm} + dm_{diff}$}
\If{$c_{dm} > k_{dm} - dm_{diff}$}
\State $sep \leftarrow calcSep(k_{i},c{i})$
\If{$sep < \theta$}
\State $possibleMatch(k_{i},c{i})$
\EndIf
\EndIf
\EndIf
\EndIf
\EndIf
\EndFor
\EndProcedure
\end{algorithmic}
\end{algorithm}     

\begin{table*}[htp]
\centering
\begin{tabular}{|c|c|c|c|c|c|c|}
\hline
 Machine & Instances  & Instance Type & CPU (equivalent) & ECUs & RAM (GB) & Cores  \\ \hline\hline
Zookeeper & 1 & t2.micro & 1 x 2.5 GHz Intel Xeon & variable & 1  & 1 \\
Nimbus & 1 & m4.xlarge & 1 x 2.4 GHz Intel Xeon E5-2676v3 & 13 &16  & 4 \\
Workers & 4 & c4.2xlarge & 1 x 2.9 GHz Intel Xeon E5-2666v3 & 31&16  & 8 \\
Workers & 8 & m4.xlarge & 1 x 2.4 GHz Intel Xeon E5-2676v3 & 13 &16  & 4 \\\hline\hline
TOTAL & 14 &- & - & 241 & 209 & 69 \\\hline
\end{tabular}
\caption{Summary of the cloud instances deployed to AWS. Here an ECU is an elastic compute unit.}
\label{tab:cloud_machines}
\end{table*}
\subsection{On-line Alert Generation}
Prototype alert generation bolts have limited functionality. These do not generate genuine alerts, since alerts are not required for our prototyping efforts. Thus alert nodes simply act as processing units, that slow down computation as though alerts were being generated. Similarly as no archival system exists, archival is not simulated.  
\subsection{Auditing}
The prototype incorporates processing nodes able to audit runtime/filtering performance. Auditing is accomplished in two ways. Per node auditing records the filtering accuracy and runtime performance, for a specific node only. This incurs a small computational overhead impacting the node being audited. As this type of auditing is only intended for use during development, it doesn't affect runtime performance during scale testing. \newline

End-to-end auditing measures filtering accuracy and runtime performance across the pipeline. This is achieved without incurring additional runtime overheads via the use of time stamps. Unique time stamps are attached to each tuple upon entering and exiting the topology. The time stamps accompany tuples through the topology, and record time to microsecond resolution. By determining the difference between time stamps, an estimation of the time taken for an individual tuple to move through the topology can be determined. Individual tuple transfer times can also be aggregated. By averaging over all tuples reaching the end of the topology, average tuple processing times can be computed. Additional metrics are also monitored at the auditing bolts. 
\section{Simulations}
Two forms of simulation were undertaken to test the prototype. The first consisted of small-scale simulations executed on a single machine. These were useful for testing and debugging the topology design and processing code. The second involved a larger scale deployment of the topology to the Amazon Elastic Compute Cloud (EC2). The cloud simulations were intended to assess the scalability of the system, and determine ease of deployment. In both scenarios the goal was to recreate the delivery of data from the SP to DP, during a plausible pulsar search scenario.
\subsection{Local `Cluster' Mode}\label{subsec:local_cluster}
Simulations were undertaken on single a computer running OSX 10.9. It possessed a single 2.2 GHz Quad Core mobile Intel Core i7-2720QM Processor, with a peak theoretical performance of 70.4 GFLOPs \citep{intel:2013:i7}. It was equipped with 16 GB of DDR3 RAM, and two 512GB Crucial MX100 solid state drives. A Storm cluster (version 0.95) was deployed on this machine, and the DP topology run in local cluster mode. This enabled testing of the framework prior to a larger scale deployment.
\subsection{Cloud Infrastructure}
We were awarded compute time upon Amazon's cloud infrastructure via the SKAO-AWS AstroCompute grant programme\footnote{\url{https://aws.amazon.com/blogs/aws/new-astrocompute-in-the-cloud-grants-program}.}. This time was used to test the performance and behaviour of the prototype, when scaled beyond a single machine. Using the Amazon Web Services (AWS) console, we provisioned a number of EC2 instances\footnote{These are virtual machines.}. The provisioned instances are described in Table \ref{tab:cloud_machines}. Note it is difficult to estimate the overall compute capacity possessed by these cloud resources. This is because EC2 instances are deployed on shared hardware, subject to load balancing policies and stress from other EC2 users. Amazon describes the compute capacity of its virtual instances in terms of EC2 Compute Units (ECUs). According to Amazon's documentation, a single ECU corresponds to a 1.0 - 1.2 GHz 2007 Intel Xeon Processor. To map this ECU unit to a meaningful value, consider the `slowest' (lowest clock speed) Xeon available in 2007, the Xeon E7310. This CPU possesses 4 cores, performs 4 operations per cycle, and has a clock speed of 1.6 GHz. To estimate the FLOPs capability of this processor, we use the formula,
\begin{gather}
\textrm{FLOPs} = \textrm{sockets} \cdot \frac{\textrm{cores}}{\textrm{sockets}} \cdot \textrm{clock} \cdot \frac{\textrm{operations}}{\textrm{cycle}} \textrm{.}
\end{gather}
The Xeon E7310 (1 ECU) is capable of a theoretical throughput of approximately 25.6 GFLOPs, according to both Equation 10 and Intel's own export specifications\footnote{See \url{http://www.intel.com/content/dam/support/us/en/documents/processors/xeon/sb/xeon_7300.pdf}.}. The 241 ECUs used during cloud experimentation, therefore correspond to an approximate computational capacity of 6.2 TFLOPs.
\section{Evaluation}
\subsection{Runtime Performance}
Topology performance is measured using the auditing nodes. These maintain statistics which are updated after each tuple is processed. Crucially, performance results differ according to the topology configuration used. The configuration describes the number of spouts and bolts in each layer, and the cardinality relationship between them. We adopt a colon delimited notation to describe the configuration, e.g. $1^{1...*}:2^{1...1}:1$. This corresponds to one input spout in layer 1, two processing bolts in layer 2, and a lone bolt in layer 3. The superscript defines the cardinality relation between the current layer and the next. For this example, there is a one-to-many relation between the spout and the bolts in layer 2, and a many-to-one relation between the bolts in layer 2, and the lone bolt in layer 3.
\subsection{Filtering Accuracy}
The spouts ensure ground truth class labels are known for all candidates entering the topology \textit{a priori}. It is therefore possible to evaluate filtering decisions at any bolt or spout. There are four outcomes for a binary filtering decision, where pulsars are considered positive (+), and non-pulsars negative (-). A \textbf{true} negative/positive, is a negative/positive candidate \textbf{correctly} filtered. Whilst a \textbf{false} negative/positive, is a negative/positive candidate \textbf{incorrectly} filtered. It is desirable to have as few false negative/positive outcomes as possible. The outcomes are eventually evaluated using standard metrics such as recall, precision, accuracy and F1 score. These are borrowed from machine learning research\footnote{We do not use imbalanced metrics here (e.g. the G-Mean or Mathews Correlation Coefficient) as we are not trying to show superior imbalanced class performance, but rather pipeline feasibility).} \citep[metrics used listed in][]{Lyon:2016:bs}. Note that we do not compare classification results obtained with the GH-VFDT against existing off-line pulsar classifiers. This is because it is inappropriate to compare off-line and on-line methods in this way - they are intended for fundamentally different processing paradigms.\newline

For the evaluation we use imbalance ratios much higher than those typically observed in the real-world. We do this for a simple reason - if pulsar examples are rare in our test data streams, the prototype will only experience minimal computational load following the classification step (i.e. most candidates will be filtered out here). Since known source matching is a relatively expensive computational operation, we wish to push the prototype to match many more candidates. Hence why we use $c_{\rm ratio} = 0.05$ or $c_{\rm ratio} = 0.1$, versus the real-world $c_{\rm ratio} = 0.0001$ (or worse).\newline

\begin{table*}[]
\centering
\begin{tabular}{|c|c|c|c|c|c|}
\hline
$c_{\rm obs}$ & Acc. & Recall & F1 & $t_{avg}$(ms) & $t_{tot}$ (s)  \\
\hline\hline
100,000 & .999 & .811 & .771 & 44 & 38 \\
200,000 & .999 & .811 & .771 & 23 & 60\\
500,000 & .999 & .811 & .771 & 59 & 80 \\
1,000,000 & .999 & .810 & .770 & 58 & 120\\
1,500,000 & .999 & .811 & .771 & 100 & 225\\
\hline
\end{tabular}
\caption{Performance \& filtering results for the pipeline run on a local `cluster'. Here  $c_{\rm obs}$ is the total number of candidates entering the topology, $t_{avg}$(ms) the time taken on average to process a tuple, and $t_{tot}$ (s) the total time for the pipeline to process all candidates. For these tests a candidate ratio of $c_{\rm ratio} = 0.05$ was used. See Section 8. for details on how to interpret the configuration used, and Section 8.1 for why such imbalance ratios were used. All results here where obtained for the configuration $1^{1...*}:4^{1...*}:11^{1...*}:4^{1...*}:4^{1...*}:4^{1...*}:4^{1...1}:1$.}
\label{tab:topology_results}
\end{table*}
\begin{table*}[]
\centering
\begin{tabular}{|c|c|c|c|c|}
\hline
Configuration & Bolts & $t_{avg}$ (ms) & Workers & ECUs  \\
\hline\hline
$2^{1...*}:2^{1...*}:11^{1...*}:2^{1...*}:2^{1...*}:2^{1...*}:2^{1...1}:2$ & 23 & 16.503 & 1 & ~20 \\
$2^{1...*}:4^{1...*}:11^{1...*}:4^{1...*}:4^{1...*}:4^{1...*}:4^{1...1}:4$ & 35 & 21.622 & 2 & ~40 \\
$2^{1...*}:8^{1...*}:11^{1...*}:8^{1...*}:8^{1...*}:8^{1...*}:8^{1...1}:8$ & 59 & 7.801 & 4 & ~80 \\
$2^{1...*}:16^{1...*}:11^{1...*}:16^{1...*}:16^{1...*}:16^{1...*}:16^{1...1}:16$ &107 & 6.891 & 8 & ~160 \\
$2^{1...*}:24^{1...*}:11^{1...*}:24^{1...*}:24^{1...*}:24^{1...*}:24^{1...1}:24$ & 155 & 2.045 & 12 & ~240 \\
\hline
\end{tabular}
\caption{Performance and filtering accuracy results for the prototype pipeline run on a remote AWS cluster. Here  $t_{avg}$ (ms) is the total time taken on average to process a tuple within the bolts (not counting communication overheads). For these tests a candidate ratio of $c_{\rm ratio} = 0.1$ was used. See Section 8.1 for details on how to interpret the configuration used, and Section 8.2 for why such imbalance ratios were used.}
\label{tab:cloud_topology_results}
\end{table*}
\section{Results}
\subsection{Local `Cluster'}
The results of local cluster experimentation are given in Tables \ref{tab:topology_results} and \ref{tab:bolts_results}. The prototype is able to process 1.5 million candidates in well under 600 seconds, as shown in Table \ref{tab:topology_results}. It is able to achieve this at a very fast rate, exceeding 1,000 candidates processed per second (up to 6,000 per second during high throughput tests). Thus the prototype is capable of running functionally at SKA scales, on present day commodity hardware. It is worth noting this result was achieved on a single laptop, possessing only 70.4 GFLOPs \citep{intel:2013:i7} of computational capacity (theoretical max)\footnote{In practice the theoretical max cannot be achieved. After taking into account operating system and Storm overheads, the true computational power available for processing is less than 70.4 GFLOPs.}. This result appears valid, as pre-processing and feature generation (the most computationally intensive tasks) require only 19 MFLOPs of compute.\newline

The prototype worked well functionally in local mode. The filtering accuracy was very high. It successfully removed most uninteresting candidates and duplicate detections. A 99\% end-to-end accuracy level is achieved using our new global sift algorithm, with a corresponding high level of pulsar recall (81\%). At the bolt level, the performance of global sift is also good, but not exceptional. Mistakes made by global sift account for why the pulsar recall rate did not move beyond 81\%. During testing global sift returned 133 out of 222 unique pulsars injected. For a trivial algorithm which can be easily modified, this result is still promising. To illustrate this, consider the accuracy achieved when using a random sifting approach. Here a 50:50 sift is employed, that propagates 50\% of tuples chosen at random. The accuracy achieved using a random sift drops to 84\% whilst recall tends to zero.\newline

Overall filtering accuracy is good, though below that achieved by recent ML methods used for candidate filtering \citep[see][]{Lyon:2016:bs,LyonPhD:1}. This result is unsurprising and excepted, since these methods are working with off-line data (models built off-line using large training sets). Yet we have still advanced the state-of-the-art, as these existing methods cannot be applied within the on-line scenarios we are considering. In these scenarios there is no training data available upfront, and severe computational restrictions.\newline

Whilst the prototype's recall rate is high, it is not high enough for use with the SKA. Applied as is, it would miss almost 20\% of all pulsars entering the system. However with improvements to sifting and known source matching possible, and the application of more sophisticated ML methods, this result can be greatly improved upon.\newline

The individual bolts in the topology appear to be runtime efficient. Each tuple required on average only a few milliseconds of processing time at each bolt. The runtime per tuple does increase as more data is processed as shown in Table \ref{tab:topology_results}. The same is true of the total runtime. There is a disparity between the average processing times shown in Table \ref{tab:topology_results}, and the bolt processing times in Table \ref{tab:bolts_results}. The disparity is accounted for by the time taken to transmit each tuple through the network. The values in Table \ref{tab:topology_results} include this transmission time, whilst the values in Table \ref{tab:bolts_results} account for only the processing. The difference serves as a reminder that transmission time plays a significant role in the total runtime of a Storm-like system. It also emphasises the importance of running tests outside of local cluster mode, which does not incur the communication overheads experienced in the real-world.

\begin{table}[]
\centering
\begin{tabular}{|l|c|c|c|c|}
\hline
Bolt & Acc. & Recall & F1 & $t_{avg}$(ms) \\
\hline\hline
Pre-processing & - & - & - & 1  \\
Global Sift & .991 & .599 & .371 & 2 \\
Feature Extraction& - & - & - & 2 \\
ML Classification& .844 & .880 & .358 & 2  \\
Source matching & 0.999 & 0.995 & .999 & 7  \\
Alert generation & - & - & - & 1  \\
\hline
\end{tabular}
\caption{The runtime and filtering performance of individual bolts. Accuracy values are only provided for those bolts filtering the data. Here $c_{\rm obs}$ is the total number of candidates entering the bolt, and $t_{avg}$(ms) the time taken on average to process a tuple. Experiments run using $c_{\rm obs}=50,000$, and $c_{\rm ratio} = 0.05$ (produces many duplicates for sifting).} 
\label{tab:bolts_results}
\end{table}
\subsection{Cloud Infrastructure}
Cloud infrastructure simulation results are presented in Tables \ref{tab:cloud_topology_results} and \ref{tab:aws_bolts_results}. Table \ref{tab:cloud_topology_results} shows topology performance according to the \textit{total time} it takes to process a tuple on average. The results in Table \ref{tab:cloud_topology_results} show an increase in processing time, when initially beginning to scale the topology (when using 1-4 workers). This is followed by a decrease in processing time when more workers are added (increasing the computational power available during execution). Note we do not show filtering accuracy results in these tables. This is because filtering performance does not change between local and remote mode - the bolts and their filters are unchanged. The only difference is that remote mode allows the topology to be scaled to utilise greater computational resources, speeding up processing times and increasing capacity. \newline

The initial increase in processing time observed when beginning to scale the topology is explainable by the results shown in Table \ref{tab:aws_bolts_results}. These describe the performance of the processing layers, as they are scaled-up (more bolts added) to process the data. When a topology contains only a few instances of each type of bolt, layers reach their processing capacity quickly. This causes an overall increase in processing times due to resource contention. As more bolts are added to the topology, counter-intuitively, the contention is not necessarily reduced. Whilst some bolts will begin to cope with the load when more instances are added, others will not. This happens when resource contention is shifted to another location in the topology, increasing processing times there instead. Note that these results were achieved when running the random 50:50 sifting approach described in Section 9.1. The random sifting approach had to be used instead of our global sift, as global sift is efficient enough to prevent us from stressing the downstream bolts significantly. As random sift propagates far more tuples, it allows us to study the scalability of the topology under more extreme loads. \newline

Contention arises when bolts undertake different proportions of the total computational workload. In the case of the pulsar search topology, known source matching is the most computationally expensive procedure (see Table \ref{tab:aws_bolts_results}). Adding more bolts to the lower layers of the pulsar search topology, increases the tuple throughput reaching known source matching bolts. Initially the matching bolts cannot meet this demand. Only when at least 8 worker nodes are available, with enough known source matching bolts, does the contention disappear and processing time decrease. This is an important observation. The computational demands of known source matching are surprising.\newline    

Table \ref{tab:aws_bolts_results} shows which bolts experienced most resource contention. This is indicated by the capacity value (CV),
\begin{eqnarray}
\textrm{CV} = \frac{(\textrm{tuples executed} \times \textrm{avg. execute latency})}{\textrm{measurement time}} \textrm{.}
 \end{eqnarray}
A value of 1.0 corresponds to a bolt at full capacity. Values greater than 1.0 indicate a bolt over capacity, and less than 1.0 under capacity. When only two bolts are present in each layer of the topology, all bolts are at or near capacity. As the topology is scaled, most bolts begin to become under utilised. The exceptions are the ML classification bolts, and the known source matching bolts.\newline

\begin{table*}[htp]
\centering
\begin{tabular}{|c|c|c|c|c|}
\hline
Bolt & Instances & Capacity & Execute Latency (ms) & Process latency (ms) \\
\hline\hline
Preprocessing & 2 & 1.984 & 1.471 & 1.955   \\
Preprocessing & 4 & 0.048 & 0.340 & 0.238   \\
Preprocessing & 8 & 0.012 & 0.167 & 0.698  \\
Preprocessing & 16 & 0.008 & 0.169 & 0.111  \\
Preprocessing & 24 & 0.008 & 0.189 & 0.369  \\\hline
Sift & 11 & 0.039 & 0.204 & 0.094 \\\hline
Feature Extraction & 2 & 1.036 & 1.570 & 1.488  \\
Feature Extraction & 4 & 0.006 & 0.176 & 0.942  \\
Feature Extraction & 8 & 0.006 & 0.141 & 0.103  \\
Feature Extraction & 16 & 0.029 & 0.620 & 0.205  \\
Feature Extraction & 24 & 0.126 & 0.595 & 0.128  \\\hline
ML Classification & 2 & 1.730 & 2.695 & 1.669   \\
ML Classification & 4 & 0.002 & 0.061 & 6.781   \\
ML Classification & 8 & 1.445 & 8.884 & 0.576   \\
ML Classification & 16 & 0.003 & 0.120 & 0.694   \\
ML Classification & 24 & 0.074 & 0.429 & 0.213   \\\hline
Known Source matching  & 2 & 0.967 & 1.846 & 1.844   \\
Known Source matching  & 4 & 0.298 & 21.167 & 6.636   \\
Known Source matching  & 8 & 0.700 & 14.232 & 16.097   \\
Known Source matching  & 16 & 0.271 & 8.148 & 5.065   \\
Known Source matching  & 24 & 0.277 & 4.829 & 4.961   \\\hline
\end{tabular}
\caption{The runtime performance of individual bolts using random sift (randomly make sift decision with 50:50 split). Random sift was used to place greater processing load upon downstream bolts in the topology. Here process latency is the time taken to `ack' a tuple after it is received. Note that `acking' a tuple involves sending an acknowledgement to the transmitter of a tuple, so it knows it has been received. Execute latency is the time taken for the bolt to complete processing a tuple. Experiments run using $c_{\rm ratio} = 0.1$.}
\label{tab:aws_bolts_results}
\end{table*}

As more bolt instances are added to each layer of the topology, execution time latency reduces, and total runtime decreases. In some cases the runtime decrease scaled linearly with the number of worker nodes used. However, this is not always the case. There are fluctuations in the results which make it difficult to discern a genuine trend. The overall results shown in Table \ref{tab:cloud_topology_results} indicate an improvement in performance scaling sub-linearly with the total number of worker nodes. This impression is based on averaged results, and agrees well with empirical experience. It is therefore likely the most accurate indication of true system scalability. We do not report total runtime, as all tests completed well within the $t_{obs}=600$ seconds required in the pulsar domain. This suggest our topology can process data in in real-time. 
\section{Conclusions}
A prototype data processing pipeline for pulsar search has been developed. It is capable of on-line and real-time operation. It employs a combination of resource efficient algorithms and optimised selection methods. Together these enable large numbers of pulsar candidates to be filtered very accurately, using limited computational resources.\newline

The performance of the prototype was first assessed on a single commodity laptop. During testing it was able to process 1.5 million pulsar candidates\footnote{The quantity delivered per SKA observation.}, in under 600 seconds. It is therefore functionally capable of processing data fast enough to meet SKA design requirements (exceeds a processing rate of 6,000 candidates per second). The prototype was also deployed to a cloud-based software infrastructure. Here the system was similarly able to process data at SKA scales, using modest computational resources (6.2 TFLOPs of processing power). The runtime performance of the prototype scaled sub-linearly with the number of worker nodes used to execute the processing. Better scaling is likely impeded by the overhead of inter-node communication (i.e. latency and bandwidth restrictions incurred due to network communication).\newline

However, the prototype was designed to favour computational efficiency over filtering accuracy and pulsar recall. This was done to ensure feasible operation at SKA scales, assuming a worst case processing scenario. This trade-off reduced pulsar recall. Thus although filtering accuracy reached 99\%, the corresponding pulsar recall rate ranged between 81-88\%. This is below the level required for SKA use. There is room to reverse the efficiency-accuracy trade-off, without significantly compromising runtime performance. It would be reasonable to double the resource use of the system, to achieve higher filtering accuracy and pulsar recall.\newline


Finally we emphasise that it was not our intention to suggest we should use so few resources (i.e. a modest ~6.2 TFLOPs easily surpassed by a single modern GPU\footnote{A NVIDIA Tesla V100 achieves 7.8 TFLOPs (double precision). See \url{https://www.nvidia.com/en-us/data-center/tesla-v100/}.}) to run a pulsar search pipeline. Rather we aimed to show that we can do so, using COTS tools at a low price point. A recall rate of 81-88\% is promising, but the remaining 12-20\% is incredibly difficult to isolate. As the recall rate increases, it becomes increasingly difficult to improve it further. Methods capable of isolating the last few percent will likely be sophisticated, and possibly be accompanied by higher computational runtime costs.\newline
 
We recommend that future work focus upon developing more sophisticated filters and learning algorithms, as these will likely greatly improve the pulsar recall rate. The algorithms installed at the bolts in the topology should also be studied in greater detail. There is scope to improve their runtime performance, and more crucially, filtering accuracy. We are currently working on a data generator that will assist us in such an investigation. 
\section{Acknowledgements}
Experimental data was obtained by the High Time Resolution Universe Collaboration. Cloud testing was supported by the SKAO-AWS AstroCompute grant programme. We thank and acknowledge the SKA Time domain team for their support, and all our anonymous reviewers for their helpful feedback.\newline

\section*{References}



\end{document}